\documentclass{article}

\usepackage{arxiv}

\usepackage[utf8]{inputenc} 
\usepackage[T1]{fontenc}    
\usepackage{hyperref}       
\usepackage{url}            
\usepackage{booktabs}       
\usepackage{amsfonts}       
\usepackage{nicefrac}       
\usepackage{microtype}      
\usepackage{lipsum}		
\usepackage{graphicx}
\usepackage{natbib}
\usepackage{doi}
\usepackage{chemformula}

\graphicspath{ {images/} }

\title{Accessible Carbon on the Moon}

\date{April 27, 2021}	

\author{ \href{https://orcid.org/0000-0001-7244-9390}{\includegraphics[scale=0.06]{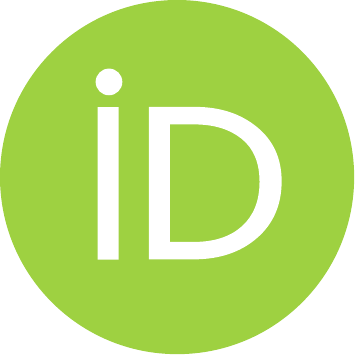}\hspace{1mm}Kevin M.~Cannon}\\
	Department of Geology and Geological Engineering\\
	\& Space Resources Program\\
	Colorado School of Mines\\
	Golden, CO 80401 \\
	\texttt{cannon@mines.edu} \\
}

\renewcommand{\shorttitle}{Carbon on the Moon}

\hypersetup{
pdftitle={Accessible Carbon on the Moon},
pdfsubject={q-bio.NC, q-bio.QM},
pdfauthor={Kevin M.~Cannon},
pdfkeywords={Moon, Carbon, ISRU, Space Resources},
}

\begin{document}
\maketitle

\begin{abstract}
	Carbon is one of the most essential elements to support a sustained human presence in space, and more immediately, several large-scale methalox-based transport systems will begin operating in the near future. This raises the question of whether indigenous carbon on the Moon is abundant and concentrated to the extent where it could be used as a viable resource including as propellant. Here, I assess potential sources of lunar carbon based on previous work focused on polar water ice. A simplified model is used to estimate the temperature-dependent Carbon Content of Ices at the lunar poles, and this is combined with remote sensing data to estimate the total amount of carbon and generate a Carbon Favorability Index that highlights promising deposits for future ground-based prospecting. Hotspots in the index maps are identified, and nearby staging areas are analyzed using quantitative models of trafficability and solar irradiance. Overall, the Moon is extremely poor in carbon sources compared to more abundant and readily accessible options at Mars. However, a handful of polar regions may contain appreciable amounts of subsurface carbon-bearing ices that could serve as a rich source in the near term, but would be easily exhausted on longer timescales. Four of those regions were found to have safe nearby staging areas with equatorial-like illumination at a modest height above the surface. Any one of these sites could yield enough C, H and O to produce propellant for hundreds of refuelings of a large spacecraft. Other potential lunar carbon sources including bulk regolith and pyroclastic glasses are less viable due to their low carbon concentrations.
\end{abstract}


\section{Introduction}

	Most studies of lunar polar volatiles have centered on water ice, including those looking ahead to harvesting resources \citep{kornuta2019, sowersdreyer2019, cannonbritt2020model}. There are a few reasons for this: (1) at least small amounts of water ice have been detected and mapped from orbit over the poles \citep{feldman1998, hayne2015, li2018}, (2) \ch{H2O} is likely the most abundant species among polar ices \citep{colaprete2010}, and (3) liquid oxygen and hydrogen (LOX+LH2) can both be derived from \ch{H2O} ice and are a widely used propellant combination. However, important future transport systems including the SpaceX Starship (Raptor engines) and others (BE-4 engines) run on liquid oxygen and methane (methalox), not LOX+LH2. In order to refuel these craft using in-space resources, a source of methane is needed; it could reduce the number of launches required if the carbon component of this methane can be found and processed offworld in a shallow gravity well like the Moon’s.

	But the Moon is not a carbon-rich world. Bulk soils from Apollo missions contained only $\sim$100 ppm levels of C from implanted solar wind, and because the Moon has no atmosphere or hydrosphere, there is no way to sequester carbon as carbonate rocks. Somewhat more promisingly, carbon-bearing molecular fragments were identified in the plume of icy material ejected by the Lunar CRater Observation and Sensing Satellite (LCROSS) experiment \citep{colaprete2010} near the Moon’s south pole. These included CO, \ch{C2H4}, \ch{CO2}, \ch{CH3OH}, and \ch{CH4}, making up a total of $\sim$5000 ppm elemental carbon in the regolith at the impact site. It is not clear how faithfully these fragments represented the carbon in the undisturbed ground, and LCROSS was a point measurement at only one location.

	Some work has been done to identify polar regions cold enough to host specific carbon-rich ices like \ch{CO2}, but the total carbon amount, favorability, and accessibility of potential carbon-rich deposits have not been studied. \citet{williams2019} provided new gridded polar temperature maps from the Diviner instrument, which can be used to highlight those areas that are permanently below the stability temperatures of supervolatiles like \ch{CO2} and \ch{C2H4}. \citet{hayne2019} identified several regions permanently below 55 K where \ch{CO2} ice is stable and showed that some observations could be consistent with carbon dioxide frosts, but the results were inconclusive. \citet{schorg2021} updated the \ch{CO2} cold trap regions for the south pole. \citet{sefton2019} presented evidence for doubly shadowed ultra-cold traps–which could host a variety of carbon-bearing phases–located within Amundsen crater. The Lunar Atmosphere and Dust Environment Explorer (LADEE) spacecraft identified an active methane cycle with surface adsorption and desorption \citep{hodges2016}, but this did not show where on the surface methane is being adsorbed. 

	Building on previous work on lunar polar water ice \citep{cannonbritt2020model, cannonbritt2020access, cannon2020}, the goals of this contribution are to: (1) estimate the carbon budget on the Moon, including a model for the Carbon Content of Ices (CCOI) at the lunar poles; (2) map locations where carbon-bearing volatiles may be especially concentrated using a Carbon Favorability Index (CFI); and (3) determine the accessibility and operability of CFI hotspots compared to other potential sources of lunar carbon. 

\section{Carbon Sources on the Moon}
\subsection{Solar wind implanted in regolith}

	Measurements of returned Apollo soil samples revealed volatiles–including carbon–implanted by the solar wind within the regolith column \citep{epsteintaylor1970}. The compilation of soil and regolith breccia data from \citet{haskinwarren1991} shows $\sim$1–250 ppm C, however the authors note significant issues with terrestrial contamination in some of these measurements. Still, native C contents of 100 ppm or higher may be plausible, and solar wind volatiles are expected to be reasonably uniform across the lunar surface (but see \citep{shukla2020}). Highlands regolith is $\sim$10 m thick on average, and mare regolith (covering 17\% of the Moon’s surface) is $\sim$5 m thick \citep{fawieczorek2012}. Order of magnitude estimates for solar wind implanted carbon can be derived by multiplying the total amount of surficial regolith by the range of C concentrations in Apollo soil data. This ignores contributions from megaregolith and paleoregolith layers at depth, but these are not nearly as accessible as surficial regolith. Solar wind implanted carbon could in theory be liberated by some combination of mechanical agitation and heating, although efficient separation from other elements is likely to be difficult, and enormous volumes of feedstock would be needed (see Fig. \ref{fig:fig7}).

\subsection{Pyroclastic deposits}

	Volcanic-derived dark mantle deposits (DMDs) have been mapped from lunar orbit \citep{head1974, hawke1989, weitzhead1998}, and the pyroclastic glasses believed to be present in these deposits contain elevated volatile contents \citep{saal2008}. However, the average carbon content of Apollo glass beads is only $\sim$0.5 ppm \citep{wetzel2015, rutherford2017}, and many DMDs are not spectrally consistent with glass making up a large fraction of the deposit \citep[e.g.][]{gaddis2003}. Therefore, even though DMDs are areally extensive, their low carbon content precludes them from being a viable source of C on the Moon.

\subsection{Polar ices}

	The LCROSS results showed that multiple frozen carbon species are present at the lunar poles (Table \ref{tab:table1}; \citet{colaprete2010}), at least in one location on the floor of Cabeus crater. Only certain molecules were detectable by the LCROSS shepherd spacecraft spectrometers, and other species including simple organics, clathrates, and aromatic hydrocarbons are also thermally stable in cold traps \citep{zhangpaige2009}. Reactions driven by energetic particles could lead to more complex refractory organic species that build up over time \citep{lucey2000}, and inorganic refractory phases could also be present \citep{steele2010} (Table \ref{tab:table1}).

\begin{table}[h]
	\caption{Carbon-bearing compounds detected in the LCROSS plume, and other potential carbon-phases that could exist in polar regolith.}
	\centering
	\begin{tabular}{llll}
		\toprule
		\textbf{Compound}                 & \textbf{LCROSS  (wt.\%)} & \textbf{LCROSS (ppm C)} & \textbf{Volatility temperature (K)} \\
		\midrule
		CO                                & {0.57}          & {2400}            & $18.9^a$                           \\
		\ch{C2H4}                         & {0.17}          & {1500}          & $44.6^a$                               \\
		\ch{CO2}                          & {0.12}          & {300}           & $59.5^a$                               \\
		\ch{CH3OH}                        & {0.09}          & {300}           & $100^a$                                \\
		\ch{CH4}                          & {0.04}          & {300}           & $25^b$                                 \\
		OCS                               & n.d.            & n.d.            & $46.8^b$                               \\
		HCN                               & n.d.            & n.d.            & $80.5^b$                               \\
		\ch{C7H8}                         & n.d.            & n.d.            & $87.6^b$                               \\
		CO·5.75\ch{H2O}                   & n.d.            & n.d.            & $36.5^a$                               \\
		\ch{CH4}·7\ch{H2O}                     & n.d.            & n.d.            & $46.5^a$                               \\
		Graphite, nano-diamonds, carbides & n.d.            & n.d.            & \textgreater{}1000     \\
		\bottomrule
		\multicolumn{4}{l}{\textsubscript{a}\citet{berezhnoy2012}, 2 cm depth. \textsuperscript{b}\citet{zhangpaige2009}}.            
	\end{tabular}
	\label{tab:table1}
\end{table}

\begin{figure}
\includegraphics[scale=0.5]{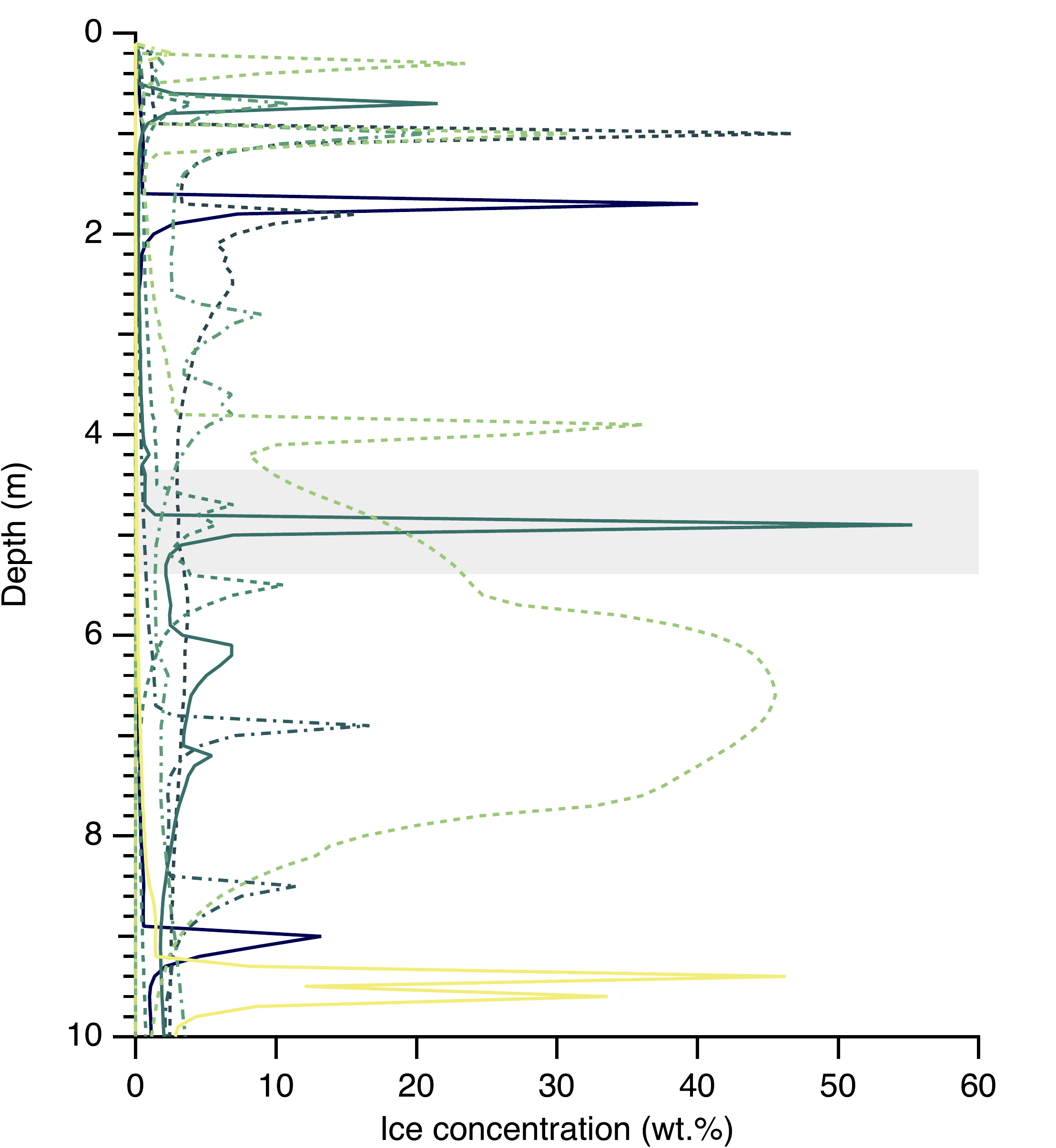}
	\centering
	\caption{Monte Carlo simulated profiles for total ice concentration as a function of depth. The gray shaded region shows a 10 cm thick horizon with an average ice content of >50\%, used in the calculations below.}
	\label{fig:fig1}
\end{figure}

	In order to extrapolate to other areas, models are needed for the amount of ice present at depth, and the CCOI. To estimate the amount of ice present in the subsurface, I used Monte Carlo simulations from a combined ice deposition and 3D regolith gardening model described in previous work \citep{cannonbritt2020model, cannon2020}. Briefly, the simulations were performed on a 500×500×Z model grid with a horizontal resolution of 1 m and a vertical resolution of 0.1 m. Ice, and ejecta from nearby craters were deposited stochastically at the surface \citep{cannon2020}, and impact gardening was simulated with individual small craters emplaced using a standard size-frequency distribution and crater production function \citep{cannonbritt2020model}. Ice loss processes \citep{farrell2019} were included as well. For this work, 10 representative subsurface ice stratigraphies were taken from the model ensembles (Fig. \ref{fig:fig1}). The stratigraphies were clipped at 10 m depth in order to focus on the more accessible near-surface materials. The true concentration of ice with depth is unknown and likely varies considerably across the lunar poles, but the columns in Fig. \ref{fig:fig1} are plausible in light of recent evidence for significant subsurface ice layers \citep{rubanenko2019, luchsinger2021}. 

	The CCOI for the LCROSS result is found by dividing the total mass of elemental carbon by the mass of all volatile species detected \citep{colaprete2010}, or $(0.49/8.61) = 5.6\%$. This metric will depend on the local thermal conditions because carbon-bearing ices will sublimate above their volatility temperatures \citep[e.g.][]{zhangpaige2009, berezhnoy2012}. To extrapolate to regions other than Cabeus, I derived three endmember volatile blends (Table \ref{tab:table2}) based on likely sources for polar volatiles including comets and volcanic outgassing. While impacts of carbonaceous asteroids may be the most important contributor to lunar ices \citep{cannon2020}, it is difficult to derive the starting mix of volatiles that would be released during vaporization from the hypervelocity impact of a C-type asteroid. \citet{berezhnoy2012} cite \ch{H2}, \ch{H2O}, CO, \ch{CO2}, \ch{H2S}, \ch{SO2}, and \ch{N2} as likely species, but the relative proportions are not available from literature sources. For comets, I used compositions from \citet{mummacharnley2011} and calculated a high-C and low-C endmember by varying their provided ranges of carbon species. The blend of volatiles from \citet{rutherford2017} was used for lunar volcanic outgassing.

\begin{table}[h]
	\caption{Volatile source endmembers used to estimate the CCOI.}
	\centering
\begin{tabular}{llll}
		\toprule
\textbf{Species} & \textbf{Comet high-C (wt.\%)} & \textbf{Comet low-C (wt.\%)} & \textbf{Volcanic outgassing (wt.\%)} \\
\midrule
CO               & 19.8                          & 0.3                          & 36.9                                 \\
\ch{CH4}              & 0.6                           & 0.2                          & –                                    \\
OCS              & 0.6                           & 0.1                          & 18.6                                 \\
\ch{H2S}              & 0.1                           & 1.2                          & 5.4                                  \\
\ch{CO2}              & 31.1                          & 2.2                          & –                                    \\
\ch{NH3}              & 0.1                           & 0.6                          & –                                    \\
\ch{SO2}              & –                             & –                            & 19.8                                 \\
\ch{CH3OH}            & 5.3                           & 0.2                          & –                                    \\
\ch{H2O}              & 42.5                          & 95.3                         & 19.3                                \\
\bottomrule
\end{tabular}
	\label{tab:table2}
\end{table}

\begin{figure}
\includegraphics[scale=0.5]{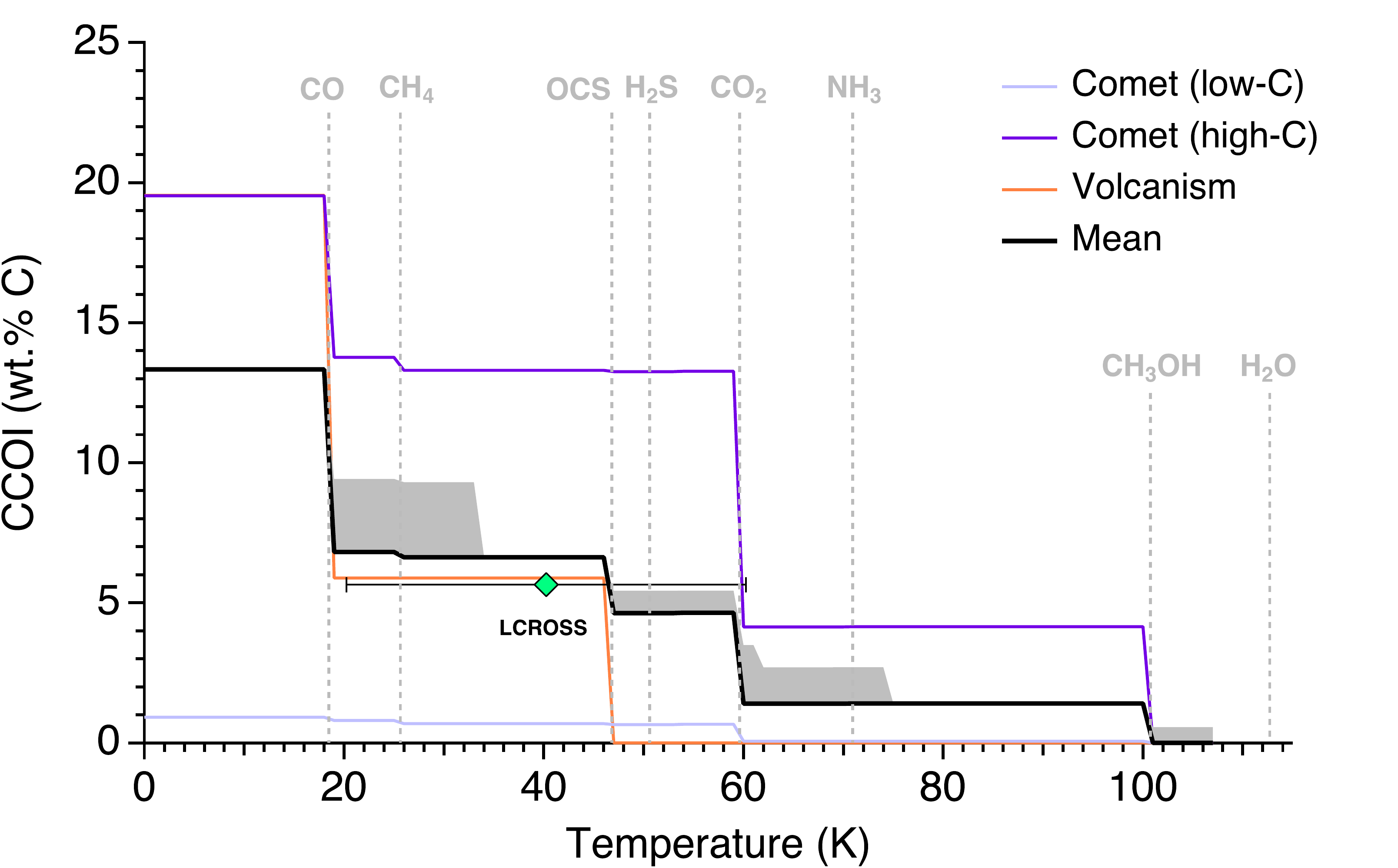}
	\centering
	\caption{Carbon Content of Ices (CCOI) as a function of temperature for three starting cases: cometary (high-C), cometary (low-C), and volcanism. Vertical lines show surface stability temperatures for specific species. The black line is the average of the three cases (used for the CFI), and the gray shaded areas represent potentially higher values in smaller ultra-cold traps.}
	\label{fig:fig2}
\end{figure}

	To calculate the CCOI as a function of temperature ($T$), the volatility temperatures ($Tvol$) from \citet{berezhnoy2012} and \citet{zhangpaige2009} were used for the species in Table \ref{tab:table2}. For a complement of carbon-bearing ice species $nc$ and total ice species $n$, The CCOI is: 
\begin{equation}
	CCOI(T) = \frac{\sum_{i=1}^{nc} w_{i,Tvol<T}fc_i}{\sum_{j=1}^{n} w_{j,Tvol<T}}
\end{equation}
	where $w$ is the mass fraction of each species, and $fc$ is the mass fraction of carbon in that compound. This implies a simplifying assumption that each volatile compound will be cold trapped at the lunar surface in proportion to its relative abundance in the gas phase. More complex atmospheric transport and chemistry models would be needed to improve on this assumption, and are left for later work. Fig. \ref{fig:fig2} shows the resulting model for the CCOI as a function of temperature. The three endmembers in Table \ref{tab:table2} are averaged together for an intermediate case that is used in the rest of the analysis below. The LCROSS value is also plotted on Fig. \ref{fig:fig2} with the minimum, mean and maximum temperatures (from Diviner) for the Cabeus impact site. The agreement with the intermediate case is quite good, but this could be partially a coincidence; the specific blend of carbon-bearing volatiles detected by LCROSS was different than any combination of those in Table \ref{tab:table2}, and there were also a significant number of chemical reactions occurring in the hot vapor plume created by LCROSS. Regardless, the intermediate case for the CCOI provides a straightforward way to estimate carbon concentrations for different temperature regimes across the lunar poles. 

\section{Carbon Favorability Index}

	Using the model for CCOI described above, I derived a CFI based on a modified version of the Ice Favorability Index (IFI) from \citet{cannonbritt2020model}. The CFI is a semi-quantitative index for locations at the lunar poles where enrichments of carbon should be found, and can be used to guide further ground-based prospecting.

\subsection{Formulation}

	The IFI is described in detail in the appendix of \citet{cannonbritt2020model}. The datasets for the CFI come from the Lunar Orbiter Laser Altimeter (LOLA; \citet{smith2010}) and the Diviner Lunar Radiometer Experiment \citep{paige2010}, both onboard the Lunar Reconnaissance Orbiter. Specifically, inputs include the Diviner maximum summer temperatures \citep{williams2019}, the LOLA topographic roughness, and the \citet{robbins2019} lunar crater database to derive a spatially resolved estimate for the crater retention ages over the poles. \citet{deutschsubmitted} found that indicators of ice at the poles seem to be correlated with maximum Diviner temperatures rather than average temperatures, hence their use here.

	The CFI has two major components: (1) a source term, and (2) a combination of carbon capture and stability. The source term is the same as in the IFI: it uses the surface crater retention age (as calculated by a moving-window approach), convolved with the time-resolved impact flux because hydrated asteroid impacts are likely the major source of polar volatiles including carbon species \citep{cannon2020}. Older terrains have higher values because they have been able to accumulate more volatiles over time. The capture and stability term is based on the temperature-dependent CCOI (described above), and surface roughness. \citet{sefton2019} suggest that smaller doubly shadowed ultra-cold traps may be $\sim$10-15 K colder than their surroundings, and could make up as much as $\sim$40\% areal fraction of traditionally defined cold traps (<112 K), with the fraction depending on surface roughness \citep{rubanenkoaharonson2017}. I make the simplifying assumption that the highest roughness values correspond to 40\% ultra-cold traps, and that this fraction decreases linearly with decreasing roughness. Taking all this together, the capture and stability term is calculated as:
\begin{equation}
	cs = (1-0.4Rn)CCOI(T_{max})+(0.4Rn)CCOI(T_{max}-15)
\end{equation}
	where $T_{max}$ is the Diviner maximum temperature, and $Rn$ is the normalized surface roughness which is calculated by taking the zscore of the surface roughness dataset, clipping between [$-2,2$] and normalizing between 0 and 1. 

\subsection{CFI results}

\begin{figure}
	\includegraphics[scale=0.1]{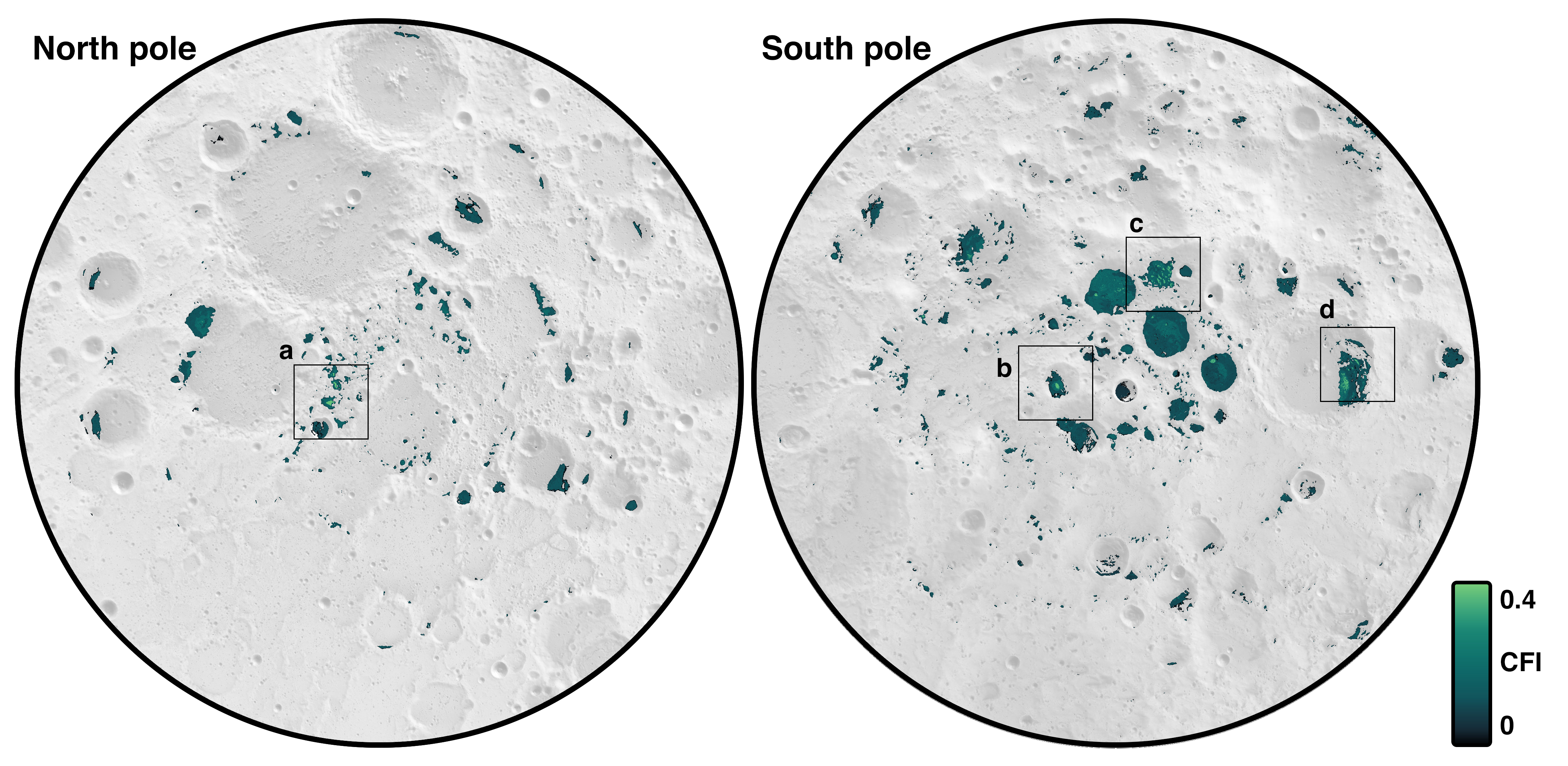}
	\centering
	\caption{Maps of the Carbon Favorability Index (CFI) for the north (left) and south (right) poles of the Moon. Boxes corresponding to CFI hotspots are shown in Fig. \ref{fig:fig4} and are described in the text. a) Intercrater Polar Highlands, b) de Gerlache crater, c) Haworth Adjacent Cold Trap, d) Amundsen crater.}
	\label{fig:fig3}
\end{figure}

\begin{figure}
\includegraphics[scale=0.25]{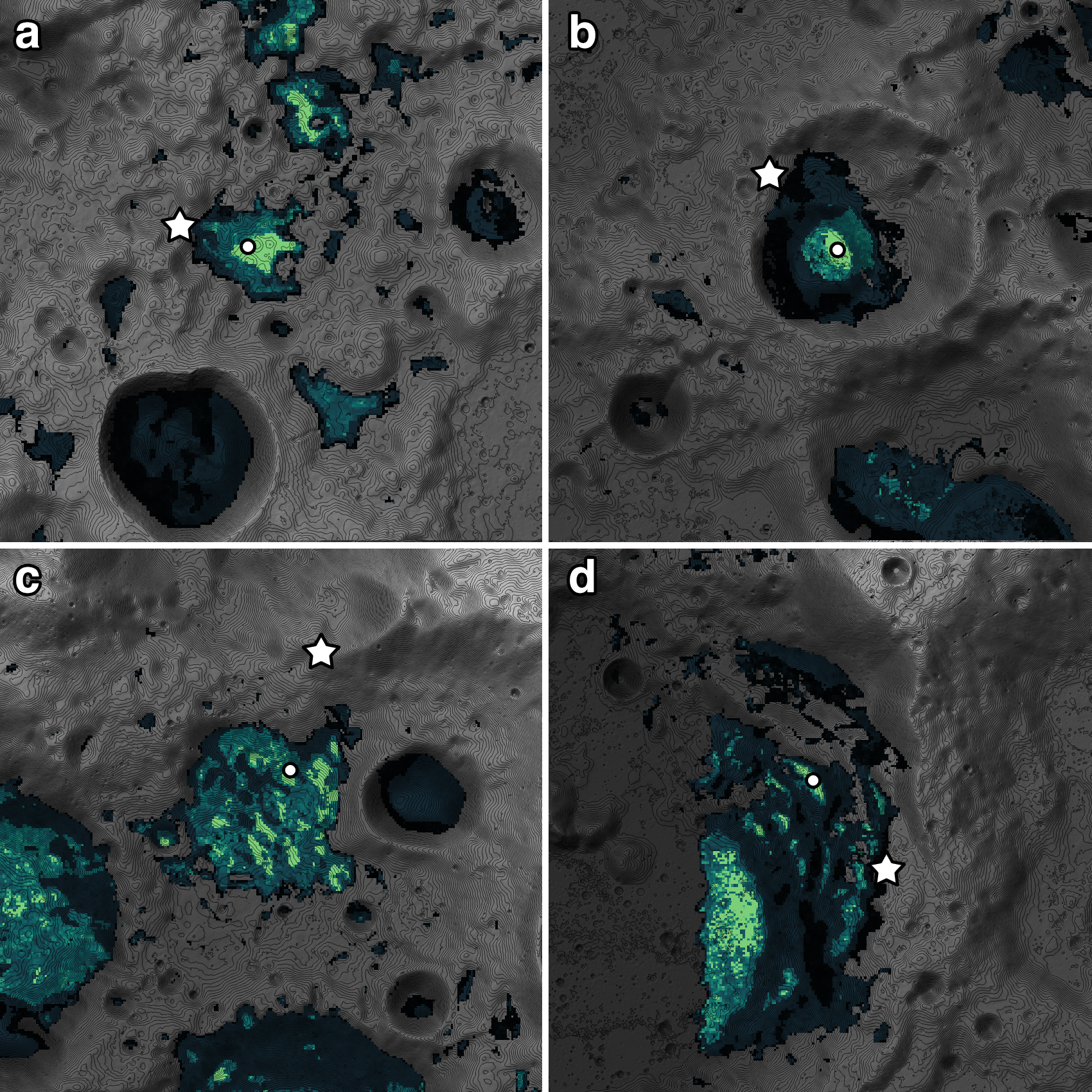}
	\centering
	\caption{Maps of CFI hotspots (boxes in Fig. \ref{fig:fig3}) and potential staging areas (white stars). a) Intercrater Polar Highlands, b) de Gerlache crater, c) Haworth Adjacent Cold Trap, d) Amundsen crater. White dots show center of CFI hotspots. Colors are the same as Fig. \ref{fig:fig3}.}
	\label{fig:fig4}
\end{figure}

	CFI maps for the north and south pole are shown in Fig. \ref{fig:fig3}. Similar to conventional cold traps, the locations with high CFI values (hereafter, “hotspots”) are smaller and more disperse at the north pole, and larger at the south pole where there happen to be a number of 50-100 km craters quite close to the geographic pole. CFI hotspots (Fig. \ref{fig:fig4}) at the north pole include several regions in the “Intercrater Polar Highlands” \citep{lemelin2014}; south pole CFI hotspots include Amundsen crater, de Gerlache crater, and an unnamed region referred to here as the Haworth Adjacent Cold Trap (HACT). The southern CFI hotspots compare favorably with calculated \ch{CO2} cold trap locations from \hbox{\citet{hayne2019}} and \citet{schorg2021}, although the CFI is based on a range of carbon species and not just carbon dioxide.
	
	Along with the CFI, an estimate for the total amount of carbon within the upper 10 m at the poles was derived using the CCOI, the Diviner maximum temperature maps, and the notional ice concentration profiles from Fig. \ref{fig:fig1}. For this calculation a single porosity of 0.2 was used, with grain densities of rock and ice of 3000 and 954 kg m\textsuperscript{-3}, respectively. The results for the 10 profiles ranged from $3.9\times10^7$ to $4.9\times10^{11}$ kg, with a mean of $1.1\times10^{11}$ kg. The significant range is due to considerable uncertainties in the subsurface ice content. 

\section{CFI Hotspot Accessibility}

	Carbon-rich deposits at the poles are not useful unless they can be readily accessed. Because of the extreme environment at the CFI hotspots (temperature, illumination, tribocharging, etc.), it likely does not make sense to deploy permanent infrastructure such as outposts or bases directly in these locations themselves. Rather, nearby staging areas (Table \ref{tab:table3}) are preferred that offer access to sun visibility (power), Earth visibility (communications), and benign terrain (permanent infrastructure, trafficability). 

\begin{table}[h]
	\caption{Details for the staging areas close to the CFI hotspots}
	\centering
\begin{tabular}{llllll}
		\toprule
\textbf{Site} & \textbf{Hotspot (lat,lon)}                           & \textbf{Staging (lat,lon)}                           & \textbf{Dist\textsuperscript{a} (km)} & \textbf{Max slope (°)} & \textbf{>1000Wm\textsuperscript{-2} (\%)} \\
\midrule
IPH           & \begin{tabular}[c]{@{}l@{}}88.534, -69.309\end{tabular}  & \begin{tabular}[c]{@{}l@{}}88.231, -71.192\end{tabular}  & 16.8                              & 18.4                   & 36.3                               \\
de Gerlache   & \begin{tabular}[c]{@{}l@{}}-88.387, -91.912\end{tabular} & \begin{tabular}[c]{@{}l@{}}-88.022, -83.075\end{tabular} & 31.8                              & 25.0                   & 43.0                               \\
HACT          & -86.586, 22.706                                             & \begin{tabular}[c]{@{}l@{}}-85.992, 20.401\end{tabular}  & 39.2                              & 24.6                   & 36.3                               \\
Amundsen      & -83.344, 84.780                                             & \begin{tabular}[c]{@{}l@{}}-83.023, 88.497\end{tabular}  & 32.4                              & 24.8                   & 48.5                              \\
\bottomrule
\multicolumn{6}{l}{\textsuperscript{a}Round-trip distance from staging area to CFI hotspot center.}
\end{tabular}
	\label{tab:table3}
\end{table}

\subsection{Access from nearby staging areas}

	Using the methods from \citet{cannonbritt2020access}, I calculated optimal paths between 4 CFI hotspots (Fig. \ref{fig:fig4}; Table \ref{tab:table3}) and nearby staging areas that had <10° slopes and >33\% average solar visibility at ground level \citep{mazarico2011}. Briefly, this involved using Dijkstra’s algorithm to find the lowest cost path between the center of the CFI hotspot, and potential staging areas that were drawn randomly from surrounding areas that met the criteria above. The cost function was based on the energy to drive a wheeled vehicle from point A to B, and takes into account both distance and topographic slope with a maximum slope set to 25°. For each of the 4 sites, staging areas were found that are within 40 km round trip distance of the CFI hotspot, and had paths with slopes <25° (Table \ref{tab:table3}).

\subsection{Solar power at staging areas}

	In the near term, solar power is the most viable power source for operations near the lunar poles. The small axial tilt of the Moon results in sunlight always coming in at a low angle near the poles, and the lunar topography causes unique illumination patterns depending on location. Solar power towers to increase illumination have been imagined for decades \citep[e.g.][]{burke1985}. At a height of 50 m, the SpaceX Starship effectively serves as its own solar power tower, and could improve duty cycles for ISRU processes compared to ground-based or shorter mast-based solar panels.

\begin{figure}[h]
	\includegraphics[scale=0.5]{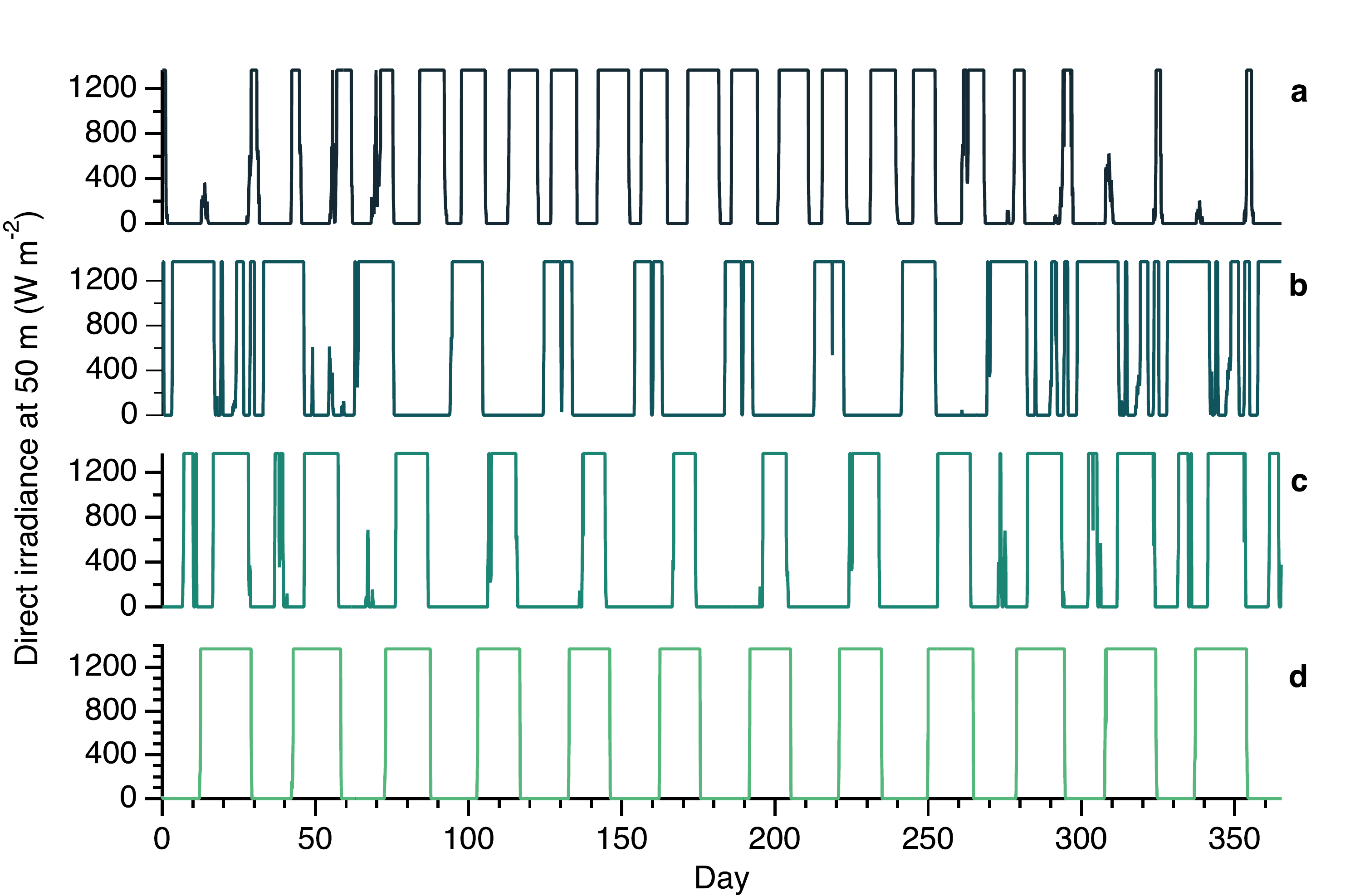}
	\centering
	\caption{Direct solar irradiance over the course of 1 year at a height of 50 m above the surface for the staging area locations in Fig. \ref{fig:fig4}. a) Intercrater Polar Highlands, b) de Gerlache crater, c) Haworth Adjacent Cold Trap, d) Amundsen crater}
	\label{fig:fig5}
\end{figure}

	An illumination model based on the horizon method \citep{mazarico2011} was used to calculate the direct solar irradiance at a height of 50 m above the surface for the 4 staging areas identified in Fig. \ref{fig:fig4}. The inputs to the model were the 80 m resolution gridded LOLA polar digital elevation models. The NASA SPICE toolkit was used for ephemeris data, and a custom solar limb-darkening model was used to convert from the fraction of the sun that is visible above the horizon to irradiance in W m\textsuperscript{-2} (limb darkening coefficient = 0.56). Models were run for 1 year starting on Jan 1, 2025. The illumination model results are shown in Fig. \ref{fig:fig5}, and irradiance metrics reported in Table \ref{tab:table3}. At two of the sites, an equatorial-like duty cycle of nearly 50\% is achieved at 50 m height, with the other two sites having lower illumination. 

\section{Discussion}

\subsection{Lunar carbon sources in context}

\begin{figure}
\includegraphics[scale=0.5]{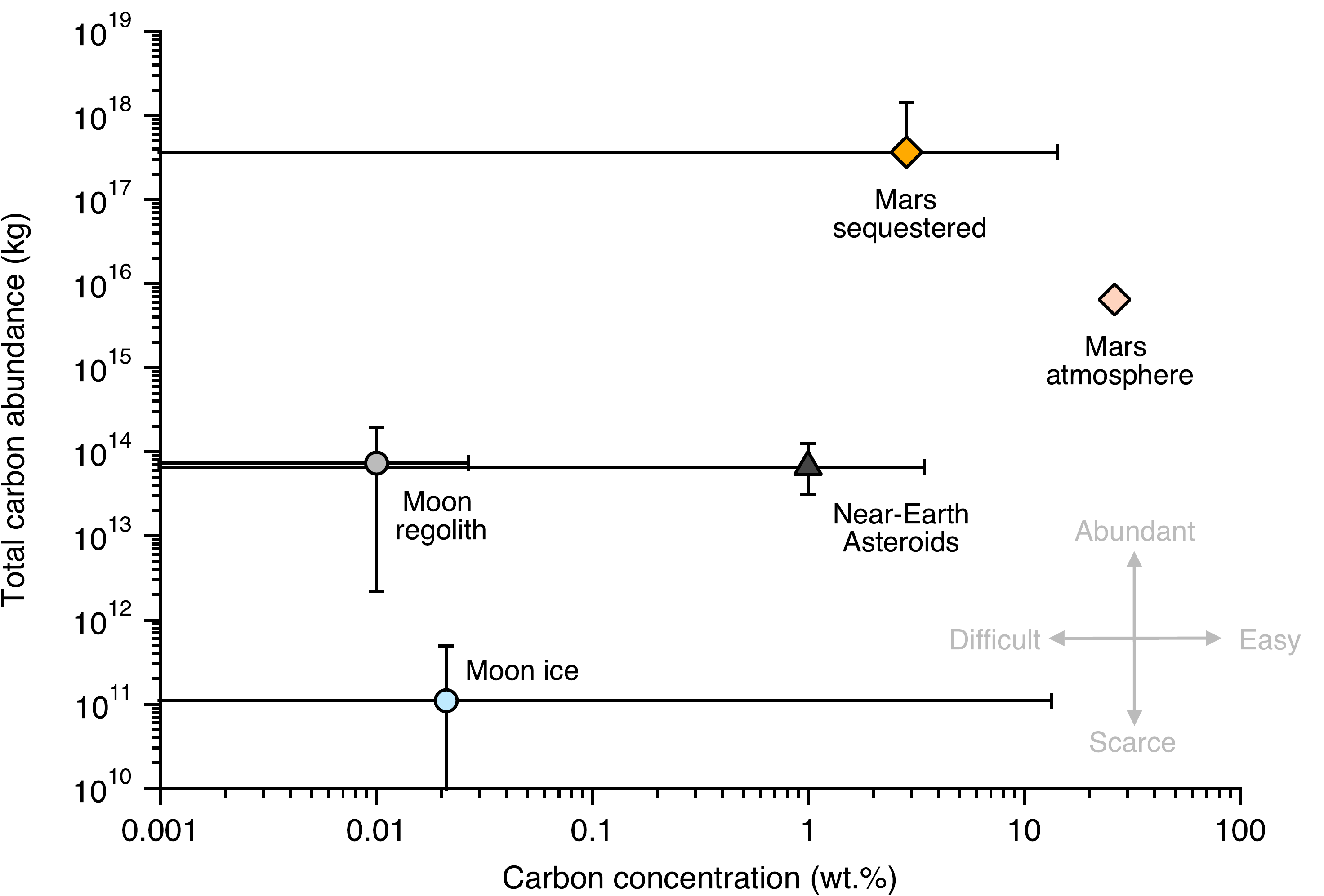}
	\centering
	\caption{Total carbon abundance versus carbon concentration ranges for inner solar system sources. Moon regolith concentrations are centered at 100 ppm, with a range from 3-266 ppm \citep{haskinwarren1991}. Abundances are found by multiplying these values by the total amount of surficial regolith (see text). Lunar ice concentrations range from 0 to the maximum CCOI value (Fig. \ref{fig:fig2}), and are centered at the mean value for the convolved CCOI and Diviner temperature maps. Total abundances are described in text. Near-Earth asteroids values assume an NEA total mass of $10^{16}$ kg, and a range of C content from 0 to 5.2\%, centered at a value that assumes 0\% C for non-C-types and 2.8\% C for C-types. Mars sequestered totals come from \citet{jakoskyedwards2018}, and are centered at a carbonate-bearing rock with 20\% magnesite (2.85\% C), with a range from from 0 to 14.25\% C (pure magnesite). }
	\label{fig:fig6}
\end{figure}

	The Moon is poor in volatiles, and it is no surprise that this includes carbon. A useful way to compare the different lunar carbon sources to each other and to other solar system objects is to plot the total amount of carbon contained within a source against the range of concentrations expected for relevant samples or localized areas (Fig. \ref{fig:fig6}). On this type of plot, potential resources trend from scarcity to abundance (bottom to top), and from difficult to easy processing (left to right). Both sources considered here–solar wind implanted carbon in regolith, and carbon ices at the poles–plot toward the lower left and are not likely to be long-term viable sources of carbon that could support a sustained human presence in space. However, the range of carbon concentrations for lunar polar ices does extend to quite high values, such that there may be low hanging fruit that are relatively easy to exploit but are also quickly exhausted.

	These values can be compared to other inner solar system carbon sources. C-type near-Earth asteroids (NEAs) are linked to the carbonaceous chondrites which contain up to $\sim$5 wt.\% C, mostly in the form of insoluble organic matter. The total amount of carbon in all the NEAs (estimated assuming the NEAs are 36\% C-types \citep{jedicke2018}, and a CM-like carbon content in C-types \citep{otting1967}) is roughly the same as the lunar implanted solar wind in the surficial regolith, although concentrations are significantly higher at least in certain asteroids. Carbon sources on Mars are much richer, with both higher concentrations and total abundances. C can be pulled from the \ch{CO2} martian atmosphere, and carbonate rocks have been located at various places on the martian surface \citep{ehlmann2008, jakoskyedwards2018}. Mars is the best source of carbon in the inner solar system, but likely pales in comparison to the amount found in Main Belt Asteroids.

\subsection{Refueling methalox spacecraft}

\begin{figure}
	\includegraphics[scale=0.2]{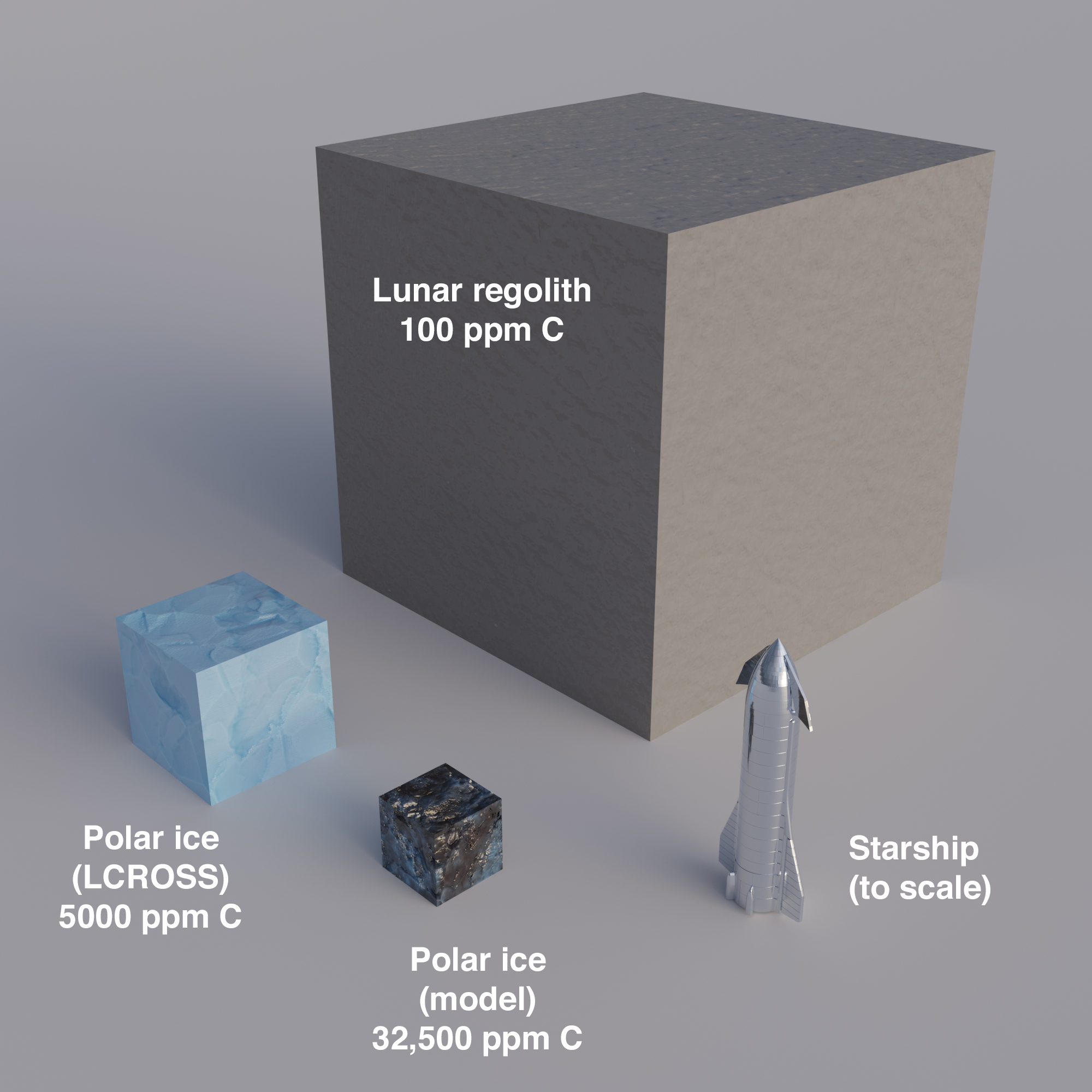}
	\centering
	\caption{The volume of material needed to produce enough carbon to refuel a SpaceX Starship (shown to scale) with 1200 tons of methalox for three different cases: bulk regolith, polar ice with the LCROSS composition, and a hypothetical carbon-bearing subsurface ice deposit with 50 wt.\% ice and 6.5\% CCOI (Figs. 1,2).}
	\label{fig:fig7}
\end{figure}

	Could the lunar polar ice deposits yield enough carbon to refuel a significant number of methalox-based spacecraft? Assuming a Starship uses 1200 tons of propellant with a 3.5:1 blend of \ch{O2}:\ch{CH4}, $\sim$200 tons of carbon are needed for each full refueling. The area of the de Gerlache CFI hotspot is about 6 km\textsuperscript{2}. If a 10 cm subsurface icy layer with 50\% ices (Fig. \ref{fig:fig1}) and 6.5\% CCOI (Fig. \ref{fig:fig2}) is present, this would yield $\sim$40,000 tons of C, which is enough for $\sim$200 refuelings. This same hypothetical layer would also yield significant amounts of O and H, such that the carbon is effectively a byproduct of oxygen and hydrogen production. The volume of required feedstock material for this modeled deposit compares favorably to the LCROSS case (8.6\% ices, 5.6\% CCOI), and of course to solar wind implanted carbon (Fig. \ref{fig:fig7}).

\section{Conclusions}

	Carbon is present on the Moon in limited amounts, mostly as solar wind implanted in the bulk regolith, and as carbon-bearing ices in the coldest regions at the poles. The solar wind carbon has a low concentration ($\sim$100 ppm), and may amount to $\sim7\times10^{13}$ kg in the total surficial regolith accessible near the lunar surface. Based on a simplified model for the Carbon Content of Ices, polar deposits may contain as much as 20 wt.\% C in the coldest regions, but are more likely to range from $\sim$0-3 wt.\% C, even in subsurface deposits that contain much more ice than the implied LCROSS results. The total abundance from all these deposits might add up to $\sim1\times10^{11}$ kg. Neither of these sources is likely to provide an abundant long-term supply of carbon to support a sustainable human presence on the Moon, and they are much less promising than martian options. However, enriched carbon hotspots that show up in the derived Carbon Favorability Index could yield enough C, H and O in subsurface icy deposits to serve as a near-term propellant source for methalox spacecraft, and are located within close distance to staging areas that provide favorable operating conditions.



\end{document}